\documentclass[conference,compsoc]{IEEEtran}
\ifCLASSOPTIONcompsoc
  \usepackage[nocompress]{cite}
\else
  \usepackage{cite}
\fi
\usepackage{graphicx}

\ifCLASSINFOpdf
\else
\fi

\usepackage{amsmath}
\usepackage{enumitem}

\usepackage{comment}
\usepackage{url}
\begin{document}
\title{On Automating Security Policies with Contemporary LLMs \\ (Short Paper)}

\author{\IEEEauthorblockN{Pablo Fernández Saura\IEEEauthorrefmark{1},
K. R. Jayaram\IEEEauthorrefmark{2}, 
Vatche Isahagian\IEEEauthorrefmark{2}, Jorge Bernal Bernabé\IEEEauthorrefmark{1}, Antonio Skarmeta\IEEEauthorrefmark{1}}
\IEEEauthorblockA{\IEEEauthorrefmark{1}University of Murcia, Spain~~ 
\IEEEauthorrefmark{2}IBM Research, USA}
}


% compsoc conferences compared to traditional conferences), use this
%Telephone: (800) 555--1212, Fax: (888) 555--1212}


\maketitle

\begin{abstract}

The complexity of modern computing environments and the growing sophistication of cyber threats necessitate a more robust, adaptive, and automated approach to security enforcement. In this paper, we present a framework leveraging large language models (LLMs) for automating attack mitigation policy compliance through an innovative combination of in-context learning and retrieval-augmented generation (RAG). We begin by describing how our system collects and manages both tool and API specifications, storing them in a vector database to enable efficient retrieval of relevant information. We then detail the architectural pipeline that first decomposes high-level mitigation policies into discrete tasks and subsequently translates each task into a set of actionable API calls. Our empirical evaluation, conducted using publicly available CTI policies in STIXv2 format and Windows API documentation, demonstrates significant improvements in precision, recall, and F1-score when employing RAG compared to a non-RAG baseline.

\end{abstract}


% \fi
\maketitle

\section{Introduction}

The increasing sophistication of cyber threats has made security policy compliance a critical component of modern software systems, spanning standalone applications, distributed architectures, and cloud-hosted services. Security policies define the necessary actions and behaviors required to protect sensitive data and maintain system integrity through three primary mechanisms: prevention, detection, and mitigation of cyber attacks. However, the growing complexity of these policies, combined with the rapidly expanding landscape of enforcement tools and technologies, creates significant challenges for automated compliance monitoring and enforcement.

Attack mitigation policies, which specify the actions required to neutralize active threats and prevent their escalation, represent a particularly challenging subset of security policies. These policies must translate high-level strategic directives into precise, executable system configurations and API calls. Traditional approaches to attack mitigation policy enforcement rely heavily on manual intervention, requiring security experts to interpret policy documents and manually configure diverse security tools. This manual process introduces several critical limitations: it is inherently error-prone due to human interpretation variability, slow to respond to rapidly evolving threats, and fails to scale effectively across large, heterogeneous computing environments.

The emergence of Large Language Models (LLMs) with advanced natural language understanding and code generation capabilities presents a promising avenue for automating attack mitigation policy enforcement. These models, trained on extensive datasets encompassing both natural language and programming code, demonstrate the ability to interpret textual descriptions and generate corresponding API calls through tool-calling mechanisms. Recent advances in LLM capabilities suggest they could potentially bridge the gap between high-level policy descriptions and low-level executable actions, enabling automated translation of attack mitigation policies into actionable system commands.

However, applying LLMs to attack mitigation policy automation presents significant technical challenges. First, the generated API calls must be both contextually accurate and precisely aligned with the intended policy objectives. Errors in policy-to-API translation could result in ineffective defenses or, worse, system misconfigurations that create new vulnerabilities. Second, attack mitigation policies often involve complex conditional logic, specialized cybersecurity terminology, and references to rapidly evolving threat landscapes. LLMs must accurately interpret these nuances while maintaining consistency across diverse tool ecosystems. Third, many organizational security tools and APIs may not be well-represented in LLM training data, particularly for custom or proprietary security infrastructure.

To address these challenges, we propose a framework that leverages Retrieval-Augmented Generation (RAG) to enhance LLM-based attack mitigation policy automation. Our approach combines in-context learning with vector-based similarity search to provide LLMs with relevant tool and API documentation during the policy translation process. This method enables accurate interpretation of high-level mitigation strategies while ensuring generated API calls align with available system capabilities.

This paper makes three key contributions. First, we present a systematic architecture for automated attack mitigation policy enforcement that decomposes high-level policies into discrete tasks and translates each task into executable API calls. Second, we demonstrate how RAG techniques can significantly improve the accuracy of LLM-generated API calls by providing contextually relevant tool documentation. Third, we provide empirical validation using publicly available Cyber Threat Intelligence (CTI) policies in STIXv2 format, showing an average 22 \% point improvement in F1-scores when employing RAG compared to non-RAG baselines across multiple LLM architectures.

The automation of attack mitigation policy enforcement has profound implications for cybersecurity resilience. By reducing response times from manual interpretation to automated execution, our approach can substantially decrease the window of vulnerability during active attacks. Furthermore, the systematic and consistent application of mitigation policies reduces human error and ensures uniform security postures across complex, distributed systems. This capability is particularly critical in environments where the speed and sophistication of attacks can overwhelm traditional manual defense mechanisms.

\section{System Design}

\begin{figure*}[htbp]
    \centering
    \includegraphics[scale=0.75]{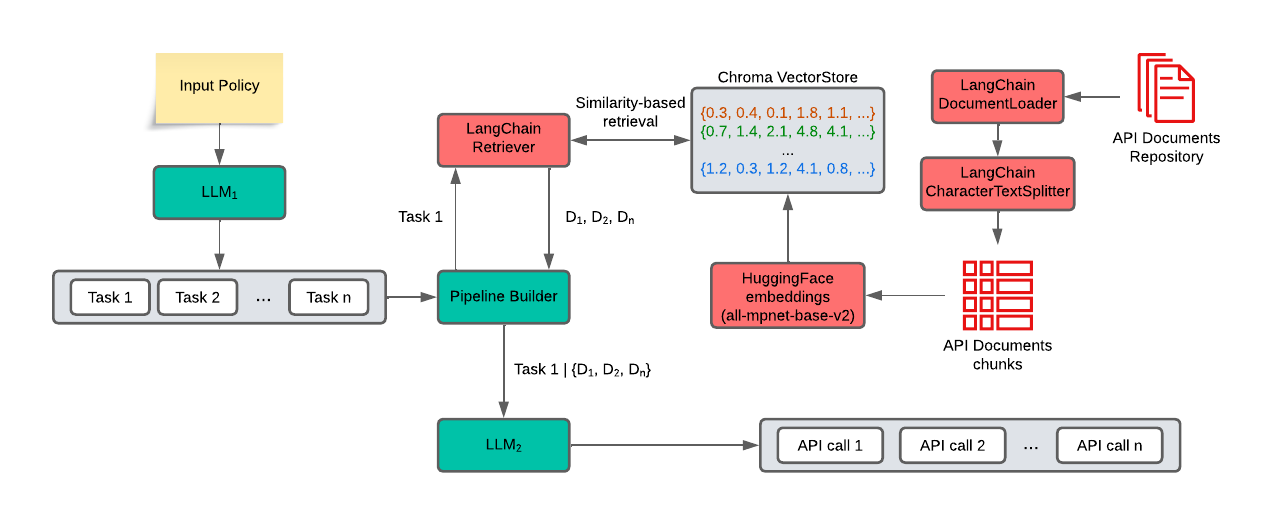}
    \caption{Proposed system architecture}
    \label{fig:arch}
\end{figure*}

We present a framework for automated attack mitigation policy enforcement that transforms high-level policy descriptions into executable API calls through a multi-stage pipeline. Figure \ref{fig:arch} illustrates our system architecture, which integrates retrieval-augmented generation with dual-LLM processing to achieve accurate policy-to-API translation. The system addresses two fundamental challenges: (1) decomposing complex policy documents into discrete, actionable tasks, and (2) translating each task into contextually appropriate API calls using relevant tool documentation.

\subsection{Tool and API Specification Management}

Our framework requires a comprehensive repository of available security tools and their corresponding API specifications. In this context, security tools encompass operating system utilities, security applications, network monitoring services, and cloud-based security platforms that can execute mitigation actions within the target environment. Each tool in the repository includes three essential components: a natural language description of the tool's security capabilities, complete API documentation specifying function signatures and parameters, and usage examples demonstrating typical invocation patterns.

The system maintains tool specifications in a structured format where each API function is documented with its purpose, required arguments, expected return values, and operational constraints. This documentation serves as the knowledge base for contextual retrieval during the API generation process. For organizational deployments, this repository can incorporate both public APIs from widely-used security tools and private APIs from custom or proprietary security infrastructure.

\subsection{Vector-Based Knowledge Retrieval}

To enable efficient and contextually relevant API retrieval, we employ a vector database approach using semantic embeddings. The system processes API documentation through a structured pipeline: first, individual API specifications are loaded using LangChain's DocumentLoader module, then segmented into coherent chunks via CharacterTextSplitter to optimize embedding quality. Each text chunk is processed through the all-mpnet-base-v2 embedding model from HuggingFace, generating dense vector representations that capture semantic relationships between API functions and their descriptions.

These embeddings are stored in a Chroma Vector Store, enabling similarity-based retrieval of relevant API documentation. During policy processing, the system queries this vector database using task descriptions as search inputs, retrieving the K most semantically similar API specifications. This approach ensures that the LLM receives contextually relevant tool documentation without being overwhelmed by irrelevant API information, addressing the challenge of limited context window capacity in current LLM architectures.

\subsection{Dual-LLM Processing Pipeline}

Our system employs a two-stage LLM processing approach to handle the distinct challenges of policy decomposition and API generation. The first LLM (LLM1) specializes in policy analysis and task decomposition, while the second LLM (LLM2) focuses on API call generation with RAG-enhanced context.

\subsubsection{Policy Decomposition Stage}
LLM1 receives the input attack mitigation policy and decomposes it into a sequence of discrete, executable tasks. This stage operates without retrieval augmentation, relying instead on carefully crafted prompts that guide the model to identify actionable components within policy descriptions. The decomposition process transforms high-level strategic directives into specific operational tasks that can be individually mapped to API calls. For example, a policy directive to ``isolate compromised network segments'' might decompose into tasks such as ``identify active network connections,'' ``disable specific network interfaces,'' and ``update firewall rules.''

\subsubsection{API Generation Stage}
LLM2 processes each task generated by the first stage, leveraging retrieval-augmented generation to produce accurate API calls. For each task, the Pipeline Builder component queries the vector database to retrieve the K most relevant API specifications based on semantic similarity. 

The retrieved API documentation, combined with the task description, forms the input context for LLM2. The model then generates a sequence of API calls that collectively accomplish the specified task. This stage employs in-context learning through carefully designed prompts that include formatting guidelines, logical ordering requirements, and representative examples of correct API call generation.

\subsection{End-to-End Processing Flow}

The complete system workflow operates as follows: Input policies are first processed by LLM1 for task decomposition without external context retrieval. The resulting task list is passed to the Pipeline Builder, which coordinates the API generation process for each task. For each task, the system queries the vector database to retrieve relevant API specifications, then provides both the task description and retrieved documentation to LLM2 for API call generation.

This architecture provides several key advantages. The separation of policy decomposition and API generation allows each LLM to specialize in its respective function, improving overall accuracy. The RAG-enhanced approach ensures that API generation operates with current, relevant tool documentation while avoiding context window limitations. The modular design enables easy integration of new tools and APIs without requiring system retraining, supporting scalability across diverse security environments.

The system's vector-based retrieval mechanism adapts automatically to new tool additions, as newly added API specifications are embedded and indexed without affecting existing functionality. This design characteristic is particularly important for dynamic security environments where new mitigation tools and techniques are continuously deployed.

\section{Experimental Methodology}

In this section, we describe the methodology used to empirically evaluate the effectiveness of our system to automate the enforcement of the attack mitigation policy through tool calls. 

%The experiment consists of several key steps, including data collection, ground-truth creation, tool specification gathering, database population, and the design of the evaluation pipeline. We provide a detailed breakdown of each step in the following subsections.

\subsection{Getting the Policies}
The first step in the experiment involved selecting a suitable set of attack mitigation policies to be processed by the system. For this, we turned to a publicly available GitHub repository maintained by MITRE, which contains thousands of Cyber Threat Intelligence (CTI) policies in the STIXv2 format \cite{mitre-cti-repo}. Structured Threat Information Expression (STIX) is a language and serialization format~\cite{stix-format} used to exchange CTI. With STIX, all aspects of suspicion, compromise, and attribution can be represented clearly with objects and descriptive relationships. 
We specifically focused on a subset of these policies that are related to attack mitigation strategies for Windows systems. These policies include a variety of security measures designed to prevent or respond to common attack scenarios on Windows-based machines. The focus on Windows-based systems was because of the continued widespread
use and because we could ensure that API descriptions of tools are available, reasonably detailed,
and vetted by a community of programmers. 

We randomly selected 10 attack mitigation policies from the repository. We had to exclude four, because they
involved steps that explicitly mentioned human participation (e.g., issue new security badges, physically remove
and lock the asset, etc.). We eventually processed six policies, which collectively contained 14 distinct tasks. 

\subsection{Creating the Ground Truth}
\label{sec:ground_truth}
The next step involved creating the ground truth to evaluate the performance of the system. In this case, ground truth refers to the manually verified sequence of API calls that correspond to each task in the policy. For this, we first manually converted each policy into a sequence of English language tasks that could be translated into API calls. For each identified task, we manually created the corresponding ground-truth API calls based on our understanding of the policy objectives and the available Windows APIs. This process involved referencing the Windows API documentation\cite{windows-api-index} and selecting the most appropriate API calls for each mitigation task. The result of this process is the dataset which will be used for evaluation, containing the mappings of each policy to a set of tasks, and each task to the correct set of API calls. 

%We divided the six policies into two groups: the first two policies, which contained six tasks, were reserved for use as in-context examples, while the remaining policies were processed and evaluated based on their translation to API calls. The ground-truth API calls served as the baseline for comparing the output generated by the system.

\subsection{Getting the API Call Descriptions}
To construct a comprehensive API knowledge base, we systematically collected Windows API documentation from Microsoft's official API reference \cite{windows-api-index}. We developed a custom web scraper using Python and BeautifulSoup to extract function names, descriptions, parameter specifications, and usage examples. This automated collection process yielded 2,637 unique API function specifications across multiple Windows subsystems including system management, network configuration, process control, and security operations.

Each collected API specification includes the function signature, natural language description of functionality, parameter details with data types, return value specifications, and relevant usage constraints. We validated the completeness of our API collection by manually verifying that all ground truth API calls were represented in the collected documentation, ensuring that our evaluation reflects realistic deployment scenarios where necessary APIs are available.

\subsection{Populating the Vector Store/Database}
With the description of the API calls collected, the next step was to populate a vector database to facilitate efficient retrieval of the relevant API calls. To achieve this, we used the LangChain library, which provides robust tools for working with language models and vector databases. Specifically, we employed LangChain’s document loaders, text splitters, and embedding models to process and store the API call descriptions.

Each API call description was loaded into the system using a document loader, and the content was split into smaller chunks using the CharacterTextSplitter. This process ensured that the text was divided into manageable segments, which could then be more easily processed by the embedding model. For this task, we used the \texttt{all-mpnet-base-v2} embedding model from HuggingFace, which is designed to generate high-quality vector embeddings for text data. The embeddings for each chunk were generated and stored in a Chroma Vector Store, which is natively supported by LangChain.

%The resulting vector database allowed the system to efficiently retrieve the most relevant API calls based on task descriptions. This database was populated with over 2600 API call entries, each of which was indexed based on its semantic similarity to other entries in the database.

\subsection{Metrics}
The final step in the experimental methodology was to execute the RAG pipeline and evaluate the performance of the system. 
Once the API calls were generated, we compared them with the manually curated ground truth using three standard machine learning metrics -- precision, recall, and F1 score, to assess how well the generated API calls matched the ground-truth API calls. We explicitly define the metrics below to avoid confusion:

\begin{equation}
\text{Precision} = \frac{\text{Number of correct API calls in output}}{\text{Total API calls in output}}
\end{equation}

\begin{equation}
\text{Recall} = \frac{\text{Number of correct API calls in output}}{\text{Total API calls in ground truth}}
\end{equation}

\begin{equation}
\text{F1-score} = 2 \cdot \frac{\text{Precision} \cdot \text{Recall}}{\text{Precision + Recall}}
\end{equation}

These metrics were calculated for each task and policy, providing a comprehensive view of the system’s ability to automate the translation of attack mitigation policies into actionable API calls. The analysis of these results is presented in the following section.

\section{Results and Analysis}
We begin by stating the initial hypotheses and objectives, followed and analysis of  the outcomes.

\subsection{Initial Hypothesis and Objectives}
The primary objective of our evaluation is to validate whether incorporating RAG improves the accuracy and relevance of the generated API calls compared to a baseline scenario without RAG. Specifically, we hypothesize that:
\begin{enumerate}
    \item The RAG-driven approach will outperform the baseline (non-RAG) for all evaluated policies.
    \item The non-RAG approach will struggle to provide correct API calls, given that the LLM has no direct background or context on the specific APIs required.
    \item Incorporating RAG will reduce the number of irrelevant or unnecessary API calls, thereby improving precision.
\end{enumerate}

\subsection{Baselines and Comparisons}
We used the \texttt{Llama-3-70b} model as the first LLM to translate the policy documents into a list of tasks. For the second LLM, which translates each task into a sequence of API calls, we evaluated the following models -- (i) \texttt{Llama-3-70b},
(ii) \texttt{Llama-3-8b} and (iii) \texttt{Mixtral-8x7b}. We did try \texttt{Llama-3-8b} and \texttt{Mixtral-8x7b} for the
first LLM but noted that the accuracy in decomposing a document into tasks was much lower -- so we exclude those results in
this paper. For each task, we computed the optimal value of \textit{K} for the similarity search, defined as the smallest number of documents recovered that includes all ground-truth API call documents. This ensures that, in the RAG scenario, the second LLM has access to all the necessary API descriptions for each task. The same \textit{K} value was used in the retrieval step for that specific task.

We compared two setups for each of the second LLMs:
\begin{description}
    \item[RAG-driven:] The LLM is provided with context in the form of the \textit{K} most similar API call descriptions retrieved from the vector database.
    \item[Non-RAG (baseline):] The LLM receives no external context on available API calls.
\end{description}

\subsection{Results}
Table~\ref{tab:example1} and Table~\ref{tab:example2} illustrate two example tasks, showing the ground truth, the maximum \textit{K} used in retrieval, the API calls returned in both RAG and non-RAG modes, and the resulting metrics. These examples were generated using \texttt{Llama-3-70b} for both the policy-to-task and task-to-API-call translations.

\begin{table*}[ht]
\centering
\begin{tabular}{l|p{1.8cm}|p{3.2cm}|p{3.2cm}|p{1.8cm}|p{1.8cm}}
\hline
\textbf{Ground Truth} & \textbf{Max K} & \textbf{API Calls Returned (RAG)} & \textbf{API Calls Returned (non-RAG)} & \textbf{Metrics (RAG)} & \textbf{Metrics (non-RAG)} \\
\hline
\begin{tabular}[htb!]{@{}l@{}}NtQuerySystemInformation\\GetLogicalDriveStrings\\QueryDosDevice\end{tabular} & 17 & \begin{tabular}[c]{@{}l@{}}GetLogicalDriveStrings\\GetLogicalDrives\\QueryDosDevice\\NtQuerySystemInformation\end{tabular} & \begin{tabular}[c]{@{}l@{}}CreateToolhelp32Snapshot\\Process32First\\Process32Next\\OpenProcess\\GetModuleFileNameEx\end{tabular} & 
\begin{tabular}[c]{@{}l@{}}Precision = 75\%\\Recall = 100\%\\F1-score = 86\%\end{tabular} & 
\begin{tabular}[c]{@{}l@{}}Precision = 0\%\\Recall = 0\%\\F1-score = 0\%\end{tabular} \\
\hline
\end{tabular} 
\caption{Example 1: Comparing RAG vs. non-RAG for a specific task.}
\label{tab:example1}
\end{table*}

\begin{table*}[ht]
\centering
\begin{tabular}{l|p{1.8cm}|p{3.2cm}|p{3.2cm}|p{2.0cm}|p{1.8cm}}
\hline
\textbf{Ground Truth} & \textbf{Max K} & \textbf{API Calls Returned (RAG)} & \textbf{API Calls Returned (non-RAG)} & \textbf{Metrics (RAG)} & \textbf{Metrics (non-RAG)} \\
\hline
\begin{tabular}[htb!]{@{}l@{}}RegOpenKeyEx\\RegEnumKeyEx\\RegEnumValue\\RegDeleteValue\\RegCloseKey\end{tabular} & 54 & \begin{tabular}[c]{@{}l@{}}RegOpenKeyEx\\RegEnumKeyEx\\RegEnumValue\\RegDeleteValue\\RegCloseKey\end{tabular} & \begin{tabular}[c]{@{}l@{}}RegOpenKeyEx\\RegEnumKeyEx\\RegQueryValueEx\\RegDeleteValue\\RegCloseKey\end{tabular} & 
\begin{tabular}[c]{@{}l@{}}Precision = 100\%\\Recall = 100\%\\F1-score = 100\%\end{tabular} & 
\begin{tabular}[c]{@{}l@{}}Precision = 80\%\\Recall = 80\%\\F1-score = 80\%\end{tabular} \\
\hline
\end{tabular}
\caption{Example 2: Comparing RAG vs. non-RAG for another task.}
\label{tab:example2}
\end{table*}

After processing all remaining tasks using each of the three second LLMs, we computed the average F1-scores for both RAG and non-RAG settings. Figure~\ref{fig:comparison} shows the comparative results across different LLMs:

\begin{figure}[htb!]
    \centering
    \includegraphics[width=0.8\linewidth]{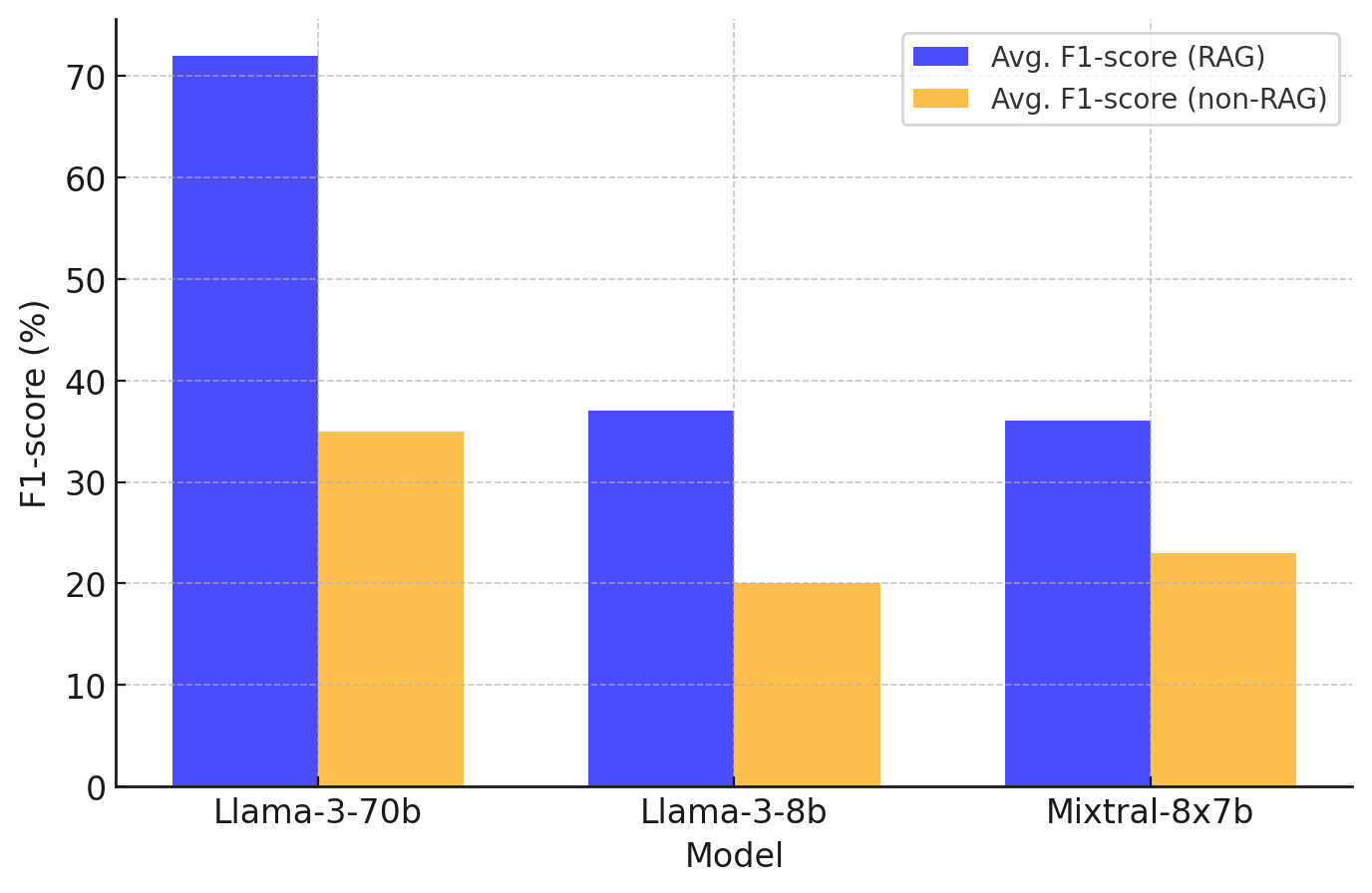}
    \caption{Comparison of average F1-scores for each model with and without RAG.}
    \label{fig:comparison}
\end{figure}

\subsection{Analysis}
The results indicate that our RAG-driven approach significantly enhances the performance of LLMs in translating attack mitigation tasks into relevant API calls. Below, we revisit the initial hypotheses:

\begin{enumerate}
    \item \textit{The RAG-driven option will outperform the baseline (non-RAG).} \\
    \textbf{Result: True.} Across all models tested, providing context via the RAG pipeline led to higher precision, recall, and F1-scores.
    
    \item \textit{The non-RAG option will struggle to provide the right API calls.} \\
    \textbf{Result: False.} Although the non-RAG approach consistently underperformed compared to RAG, it did produce decent results for certain tasks, suggesting that some Windows API data may be present in the base LLM training.
    
    \item \textit{The RAG-driven option will reduce the number of irrelevant API calls (improved precision).} \\
    \textbf{Result: True.} The precision metrics demonstrated that RAG-based queries filtered out extraneous API calls more effectively, thus aligning the output more closely with the ground truth.
\end{enumerate}

Moreover, the data show that using RAG confers an average improvement of about 22 \% point in F1-scores across all models. Notably, the largest model (\texttt{Llama-3-70b}) benefited the most from context retrieval, experiencing a 37 \% point increase in F1-score compared to its non-RAG counterpart. Interestingly, in RAG mode, smaller models match or even outperform the non-RAG version of the largest model, highlighting the potential effectiveness of this technique for the proposed use case.

%\section{Discussion}

\section{Implications for Dependable Security}

 By improving the accuracy, scalability, and responsiveness of security defenses, our approach contributes to building more resilient systems that can proactively respond to evolving cyber threats.

\begin{enumerate}

\item \textbf{Consistency:} The ability to translate high-level policies into precise API calls reduces the chances of human error and ensures that security measures are applied consistently and quickly. 

\item \textbf{New threats and policies:} One of the major challenges in contemporary cybersecurity is the rapid evolution of attack techniques and strategies. Traditional security tools often rely on signature-based methods or static rule sets, which can become obsolete when faced with new, previously unseen attack vectors. To remain effective, security systems must be able to adapt in real-time, updating their mitigation strategies as new threats arise. In this context, new policy documents can be written in response to
new threats and immediately automated by systems like ours using existing tools. 

\item \textbf{New Mitigation Tools:} On the flip side, information about
new tools, APIs, and other mitigation strategies can be added to the vector 
database without disturbing existing automations. Since the system uses RAG, 
it can select the most relevant tools and API actions based on real-time data, enabling it to easily adapt to new mitigation tools.

\item \textbf{Scalability:} The automation of attack mitigation policies provides a scalable solution for security in large and complex systems. The ability of our system to scale and handle a diverse range of tools and APIs makes it an ideal solution for securing large-scale environments, including cloud platforms, distributed networks, and IoT ecosystems. 

\end{enumerate}

\section{Related Work}

The integration of LLMs in the field of cybersecurity is emerging as a major trend in the literature. Several benefits have been identified when applying this kind of models to various key areas such as vulnerability detection, malware analysis, network intrusion detection, or to analyze and extract knowledge from high-level security artifacts such as security and privacy policies \cite{xu2024}. Moreover, it is being demonstrated that LLMs are not limited to question-answering, but are also capable of executing actions, including enforcing security measures \cite{zhang2025}.

Beyond their analytical capabilities, LLMs are useful to automate security policy management. Recent work explore the design of methods to translate security policies formulated in natural language to a machine-readable format which can be easily and effectively verified and enforced \cite{martinelli2024}. Additionally, LLMs are proposed to assess the compliance between cybersecurity controls and organizational policies, assisting in solving challenges related to efficiency or accuracy, and ensuring security measures allign with regulatory standards \cite{salman2024}.

In \cite{aghaei2022}, authors have created a domain-specific natural language model capable of identifying text connotations which are typical to the cybersecurity field. The resulting fine-tuned model outperforms the competence when evaluating the automation of many critical cybersecurity tasks. In a similar line, other research efforts have enhanced the ability of LLMs to process and analyze threat intelligence information by integrating retrieval-augmented generation (RAG) techniques, also achieving a more domain-specific model \cite{singh2024}.

Closer to our research, some studies have analyzed the automation of cybersecurity decision-making and policy enforcement using LLMs. One approach \cite{yang2024} presents a novel framework built upon LLMs to automate threat modeling in banking systems. The solution assists in mapping descriptions of the banking system design, to potential security threats, and generate mitigation strategies based on those. Another significant study \cite{jjiandong2024} involves the use of LLMs for strategic cybersecurity reasoning. They propose a system which correlates CVEs with MITRE ATT\&CK techniques, by creating a human-judged dataset used for a retrieval-aware training of the model.

While these studies propose interesting solutions and analyses of the growing research field of LLMs applied to cybersecurity, there are some research gaps that remain to be addressed. The existing literature primarily focuses on threat analysis, policy translation and decision support, but few studies propose a fully automated mechanism to enforce security policies based on high-level descriptions. Many of these concentrate on static policy generation or compliance assessment, rather than translating security policies into executable security actions. Our work advances the state-of-the-art by closing the gap between security policy generation and enforcement, leveraging and evaluating several LLMs to automatically translate security policies into specific API calls which can be directly enforced, ensuring that policies are not only analyzed but also implemented in real-world security environments.

\section{Limitations and Conclusions}

The main limitation of our study is that we focus on attack mitigation policies
for Windows systems. This is primarily a consequence of us trying to get an end-to-end implementation and workflow going. We are actively exploring other application
areas, deployment models and security policy types and evaluating the accuracy
of automating them using LLMs. Another area of active research is how to incorporate
\emph{human-in-the-loop} tasks while transforming security policy documents.

In conclusion, the automation of attack mitigation policy enforcement using LLMs and tool calling has the potential to revolutionize the way security is implemented and maintained in modern systems. By improving the dependability, scalability, and adaptability of security measures, this approach contributes to the development of more resilient systems that can proactively defend against emerging threats. Furthermore, by integrating security directly into the systems engineering process, it fosters a culture of \emph{security by design}, ensuring that security is an integral part of the software development lifecycle. As this technology continues to evolve, its impact on dependable security and systems engineering will only continue to grow, providing a robust foundation for the protection of critical infrastructure in an increasingly complex digital world.

\section{Acknowledgments}

This work has been funded by the Horizon projects CLOUDSTARS (GA: 101086248) and ResilMesh (GA: 101119681), and also by the European Union's NextGenerationEU, as part of the Recovery, Transformation, and Resilience Plan, supported by Spanish INCIBE (6G-SOC project).



%\bibliographystyle{IEEEtran}
\bibliographystyle{IEEEtran}  % Use IEEE style
\bibliography{references}     % Name of the .bib file without extension

\end{document}